\documentclass [12pt]{article}

\usepackage[T2A]{fontenc}

\usepackage[cp1251]{inputenc}
\usepackage[english]{babel}
\usepackage{amssymb}

\usepackage[dvips]{graphicx}
\usepackage{epsfig}
\usepackage{float}
\usepackage{slashed}

\usepackage{subcaption}

\newcommand{\be}{\begin{equation}}
\newcommand{\ee}{\end{equation}}
\newcommand{\bea}{\begin{eqnarray}}
\newcommand{\eea}{\end{eqnarray}}

\def\p{\partial}
\def\pslash{\p\raise.3ex \hbox{\kern-.5em /}}
\def\delslash{\nabla\raise.3ex \hbox{\kern-.7em /}}

\begin{document}
\begin{center}
{ \textbf{ {    \textbf{An upper limit on fermion mass spectrum in
non-Hermitian models and its implications for studying of dark
matter } }} }
\end{center}

\begin{center}
{ \textbf{  { \textbf{V.N.Rodionov*, A.M.Mandel**}}}}
\end{center}

\begin{center}
*{REU, Moscow, Russia,  \em E-mail    rodyvn@mail.ru}
\end{center}

\begin{center}
**{MSTU STANKIN, Moscow, Russia,  \em E-mail arkadimandel@mail.ru
}
\end{center}

\begin{center}

\abstract{  The paper formulates  a principal positions of
non-Hermitian models with $\gamma_5$-mass extensions,
 which often be ignored in  some investigations for this subject.
 In fact in this case  Hamiltonians contain not only Hermitian masses
 $m_1$, but also contribution from anti-Hermitian components of fermion
 masses $m_2$.  Main misunderstanding a number of papers is consist in using
 of this model for any values of fermion masses for  fixed values of
 $m_1$ and $m_2$.  However the basis appearing of two parameters masses
 may be undertaken a simple estimation for determination maximal
 permissible value  of fermion mass $M={m_1}^2/2m_2$ which may be
 used for this model. Easy to see that $M$  becomes infinite and hence
 experimentally doesn't observable only  in Dirac's limit, when the
 non-Hermitian mass fully is disappearing.   In  particular the
 equality $m_1=m_2$ can be realized only in two cases  when $m_1=m_2=0$
 and $m_1=m_2=2M$.  Moreover  in the second case the  question is  about
 the possibility existence a number  of new fermions
 masses which are equal to the masses of  particles Standard Model(SM)
 but when they  have the non-Hermitian characteristics.
 In this case the paradox of the "two masses" takes place and its solution
 may be done only in suggested model with a maximal mass.
 Appearing particles can be  considered  as some new particles arising  beyond SM.
 The unusual  properties of these  particles allow also to consider their as
 possible candidates  in  structure of dark matter.    }
\end{center}

\section{Introduction}

As it is now known, the defining feature of elementary particles
is that their properties and interactions may be characterized
properly in terms of local fields. The following question may be
posed in the same terms: should the mass of elementary particles
be limited from the above? To put this another way, what is the
maximal mass of a particle that still has the local field concept
applicable to it? No experiments focused on finding particles with
the maximal mass have been carried out yet. It is known only that
a top quark is the heaviest particle in the Standard Model (SM).
Naturally, the scope of experimental search for “maximons” is
limited by the feasibility to construct very high-energy
accelerators. However, a detailed study of models with a maximal
mass may reveal new and unique opportunities for detecting the
effects stemming from the mentioned limitation. This refers to
various external influences that, when taken into account, may
reveal a number of effects induced by the limited mass spectrum of
elementary particles. For example, if the interaction with intense
magnetic fields is considered, several effects may become
observable. The emergence of the so-called exotic particles, which
were predicted by V.G. Kadyshevsky in the geometric approach [1],
is one of the possible effects of finiteness of the mass spectrum.

It should be noted that the idea of finiteness of the mass
spectrum of elementary particles has been proposed first in 1965
by M.A. Markov \cite{Markov1}. This finiteness was associated with
Planck mass
 \, $M_{Plank} = \sqrt{\hbar c
/ G} \sim {10}^{19}$ GeV, where $G$ is the gravitational constant,
$\hbar$ is the Planck constant, and the $c$ is  speed of light,
and was written as \cite{Kad1} \be\label{Lfund1} m \leq M_{Plank}.
\ee

The particles with maximal mass, which were called by the author
maximons, hold a special place among elementary particles. In
particular    the concept of maximons formed the basis of the
Markov scenario of the early Universe [3]. However, condition (1)
was initially a purely phenomenological and was not taken into
account in the development of the theory.

A new radical approach involving the actual introduction of the
finiteness condition into the theory was proposed by Kadyshevsky
at the end of the 1970s [2].  Markov’s concept of the maximal
particle mass was regarded in this approach as a new fundamental
physical principle of quantum field theory (QFT). The condition of
finiteness mass spectrum  was postulated in the proposed theory in
the form:
\begin{equation}\label{Mfund1}
m \leq  M,
\end{equation}
where maximal mass parameter (fundamental mass) was considered as
a new physical constant. This quantity was regarded as the
curvature radius of a five-dimensional hyperboloid when its
surface is a realization of the curved momentum 4-space. Using an
anti-de Sitter space for this purpose, one easily obtains the
following relation:
\begin{equation}\label{O32}
    p_0^2 - p_1^2 - p_2^2 - p_3^2 + p_5 ^2 = M^2,
\end{equation}
from which it follows that at the mass surface for a free particle
${p_0}^2-\vec{p}^2=m^2$ inequality (\ref {Mfund1})is realized
automatically.

It is  important  that in the geometric model  containing
restrictions of  fermion masses, may be observed some new
particles, which were named by the "exotic fermions".  At once
there was  suggested that  arising of new fermions exclusively be
attributed  to the development of geometric representations. In
particular this fact was explained by the appearance of additional
degree of freedom which has been appearing thanks to presence of
the  fifth  component of fermion momentum $
\varepsilon=p_5/|p_5|=\pm 1$ in the anti-de Sitter space
\cite{Max}.

 However at present  it is absolute clear that appearance the exotic
 fermions it is not prerogative of  the geometric approach.
 In particular, in purely algebraic model, which contains both Hermitian
 ($m_1$) and anti-Hermitian massive components of fermions
 ($m_2$) when are provided that
 \begin{equation}\label{888} m^2= {m_1}^2 -{m_2}^2, \end{equation}
 restriction  of  fermions mass and presence of exotic particles
 also  exist \cite{Max}.  This is becoming obvious on the basis of
 theorem about relation between
 arithmetic  and geometric  averages values for two positive numbers.
 If we take into account this inequality which in expanded form
 looks like as   ${m_2}^2+ {m}^2  \geq 2\sqrt{{m_2}^2 *{m}^2}$ then
 from  it follows restriction $m\leq M=(m_1)^2/2m_2$ and if to consider
 (\ref{888}) we can obtain that "exotic fermions"
 really here also  are  appearing \cite{ROD1}.

The properties of exotic particles differ radically from the
characteristics of their well-known partners. Besides it turned
out that the geometric approach is not  single prerequisite to the
emergence of such particles in the theory. Indeed, the development
of pseudo-Hermitian $PT$-symmetric quantum theory has shown that
these particles emerge as a consequence by itself of finiteness of
the mass spectrum of elementary particles. Thus, the experimental
search for exotic particles may result in the discovery of the
maximal mass values itself. In particular this approach becomes
feasible owing to presence the calculation of the spectrum of
energies of a neutral fermion with an anomalous magnetic moment
(AMM) in  theory with a maximal mass\cite{ROD2,ROD3}.

In our approach developed in [5, 6] is used procedure  which
differs considerably from the methods  used earlier. In
particular, we obtained that the heaviest fermions (maximons with
mass $M$ ) may be a crucial component of dark matter and that they
have to consist of pseudo-fermion components. They also should
possess a modified nature  of interactions. Therefore, both
theoretical and experimental study of pseudo-Hermitian
characteristics of massive fermions assumes a particular
importance. In a result  the development of guidelines the
determination of constraints on the parameters of  maximons  one
can't improbable that new physical phenomena may be detected at an
energy of several  TeV. It should be noted that this is the
centerpiece of the research program for the Large Hadron Collider
(LHC) at CERN [7, 8]. It is already becoming widely understood
that some of the basic principles of the SM require further
refinement.  Specifically, it concerns the description of the
spectrum of fermion masses, which is not yet unlimited in the SM.
However, a very wide range of masses of known elementary particles
is found already in the Standard Model itself. For example, the
mass of a top quark (the heaviest known elementary particle) is
approximately 300000 times greater than that of an electron. All
this suggests that late or early  we simply will be forced to take
requirements of  restrictions of spectrum mass because it  will
become necessary.

Thus, the issue about neutral maximons with AMM is of considerable
interest in the context of their probable inclusion into the
structure of dark matter. Astrophysical studies may prove key
factor here, since the clarification of unusual pseudo-Hermitian
characteristics of the considered fermions is of paramount
importance in such studies. It should be noted that several
research groups throughout the world are already searching for
marks of dark matter in cosmic rays. The most ambitious project of
this kind is known as IceCube [9]. More than 5000 high-sensitivity
IceCube sensors, which are installed within the Antarctic ice
sheet at the Geographic South Pole, collect data on galactic
fluxes of various particles. These detectors are deployed at
depths ranging from ~1500 to ~2500 m. The sensors thus cover a
cubic kilometer of ice, and the researchers hope that this
galactic-ray experiment may help examine deep-space sources.

These studies will surely be instrumental in solving certain
important problems related to both low-energy neutrinos and
high-energy particles. Specifically, such experiments are
concerned with the examination of the internal structure of
supernovae,  the processes of formation of black holes and gamma
ray bursts. The current theory implies that interactions in these
objects are guarantee  accelerate particles to exceptionally high
energies. According to several astrophysical estimates, these
energies may reach ${10}^{6} - {10}^9$ GeV [9]. It is likely that
the masses of new particles, which could be regarded as dark
matter candidates, also fall within this range. Thus,
astrophysical estimates are way beyond the limit of masses probed
at LHC, where a targeted search for heavy exotic particles is
being also carried out [8].

 It is intuitively clear that studies which let
covering the widest possible energy range should yield the most
interesting results in this field and, that is no less important,
could finally shed light on the mysteries of dark matter. It
should be noted in this context that exotic particles emerging in
theories with a maximal mass is noted by individual "code"(a set
of pseudo-Hermitian characteristics), which differentiates them
from common particles found in the SM. According to our estimates,
the maximal mass may be bounded by $M = 2 *{10}^{14} GeV$ [10]; at
this level, the magnetic moment of exotic neutrinos were of the
order of ${10}^{-19}\mu_0$, where $\mu_0$ - is the Bohr magneton,
and the magnetic field was evaluated at 8000 G. However, if one
assumes that the magnetic moment of a neutrino may be somewhat
larger (see, for example, the GEMMA collaboration data [11]), the
limit on the maximal mass may enter the domain of astrophysical
observations ($M\sim {10}^8$ GeV; see [9]).

In addition, one should consider the fact that giant magnetic
fields produced by pulsars and magnetars may also be present in
experiments on neutrino astrophysics and cosmic rays. This, the
formulas for analytical and numerical evaluation of
characteristics of massive fermions propagating in magnetic fields
with an intensity of $\sim 10^{15} - 10^{16}$ G [12] should be
applicable to these enormous (relative to the scales familiar to
us) values. Note that the exact solutions for the energy of
neutral pseudo-fermions with AMM propagating in magnetic fields
[5, 6] may help cover the range of ultrahigh energies (up to $\sim
{10}^9$GeV) and intense magnetic fields (up to $\sim {10}^{16}$
G).

Thus, further development of the theory established by Kadyshevsky
may provide specific guidelines for future experiments focused on
the search for exotic fermions. Specifically, laboratory
experiments with low-energy polarized neutrinos propagating in a
magnetic field may be the least time-consuming and laborious.
Arguably, precision experiments on the measurement of mass of
neutrinos (e.g., the Troitsk tritium experiment), where weakly
excited neutrino fluxes propagate in control and fairly intense
magnetic fields, fit the requirements in this case.

It was already noted that a limited mass spectrum might be
obtained not only in the geometric approach, but also in
non-Hermitian (pseudo-Hermitian) fermion systems, which have a
direct application in neutrino physics. Systems of this kind are
called - $PT$-symmetric models and are used in various areas of
modern physics. Specifically, theoretical and experimental studies
in non-Hermitian optics started more than ten years ago.

\section{ Modified Dirac model for  non-Hermitian mass parameters}

Let us now consider the modified Dirac equations for free massive
particles using the ${\gamma_5}$-factorization of the ordinary
Klein-Gordon operator. In this case we will make similar actions
as for  known Dirac procedure. As he himself wrote: "...get
something like a square root from the equation Klein-Gordon"
\cite{Dir}, \cite{Dir1}. And really if we shall not be restricted to only
Hermitian operators then we can  represent the Klein-Gordon
operator in the form of a product of two commuting matrix
operators with $\gamma_5$-mass extension:

\be\label{D2} \Big({\partial_\mu}^2 +m^2\Big)=
\Big(i\partial_\mu\gamma^{\mu}-m_1-\gamma_5 m_2 \Big)
\Big(-i\partial_\mu\gamma^{\mu}-m_1+\gamma_5 m_2 \Big), \ee where
 the physical mass of
particles $m$ is expressed through the parameters $m_1$ and $m_2$
\be \label{012} m^2={m_1}^2- {m_2}^2. \ee

For  the function would obey to the equations of Klein-Gordon
\be\label{KG} \Big({\partial_\mu}^2
+m^2\Big)\widetilde{\psi}(x,t)=0 \ee
one can demand that it also satisfies to one of equations of the first order
\be\label{ModDir}
\Big(i\partial_\mu\gamma^{\mu}-m_1-\gamma_5 m_2
\Big)\widetilde{\psi}(x,t)
=0;\,\,\,\Big(-i\partial_\mu\gamma^{\mu}-m_1+\gamma_5 m_2 \Big)
\widetilde{\psi}(x,t)=0 \ee

Equations (\ref{ModDir}) of course, are less common than
(\ref{KG}), and although every solution of one of the equations
(\ref{ModDir}) satisfies to (\ref{KG}), reverse approval has not
designated. It is also obvious that the Hamiltonians, associated
with the equations (\ref{ModDir}), are non-Hermitian, because in
them the $\gamma_5$-dependent mass components appear ($H\neq
H^{+}$):

  \be\label{H} H =\overrightarrow{\alpha} \textbf{p}+ \beta(m_1
+\gamma_5 m_2)\ee  and \be\label{H+} H^+ =\overrightarrow{\alpha
}\textbf{p}+ \beta(m_1 -\gamma_5 m_2).\ee Here  matrices
$\alpha_i=\gamma_0\cdot\gamma_i$, $\beta=\gamma_0$,
$\gamma_5=-i\gamma_0\gamma_1\gamma_2\gamma_3$.   It is easy to see
from (\ref{012}) that the  mass $m$, appearing in the equation
(\ref{KG}) is real, when the inequality \be \label{e210}
{m_1}^2\geq {m_2}^2.\ee is accomplished, but this area contains
descriptions not only ordinary particles, but  also the the exotic
particles which do not subordinate to the ordinary Dirac equation.

In this section, we will also want touch upon question of
describing the motion of Dirac particles, if their own magnetic
moment is different from the Bohr magneton. As it was shown by
Schwinger \cite{Sc}  the equation of Dirac particles in the
external electromagnetic field $A^{ext}$ taking into account the
radiative corrections may be represented in the form: \be\label{A}
\left({\cal P}\gamma - m\right)\Psi(x)-\int{\cal
M}(x,y|A^{ext})\Psi(y)dy=0, \ee where ${\cal M}(x,y|A^{ext})$ is
the mass operator of the fermion in the external field and ${\cal
P}_\mu =p_\mu - {A^{ext}}_\mu$ . From equation (\ref{A}) by means
of expansion of the mass operator in a series of according to $
eA^{ext}$ with precision not over then linear field terms one can
obtain the modified equation. This equation preserves the
relativistic covariance and consistent with the phenomenological
equation of Pauli obtained in his early papers (see for example
\cite{TKR}).

\section{Non-relativistic limit modified Dirac
equation in the electromagnetic field with  $\gamma_5$-mass
extension.}

In a most cases no necessity in exact solutions of modified Dirac
equation with non-Hermitian $\gamma_5$-mass extensions.
Really one can confine by non-relativistic   amendments   using expansions
$\sim v/c$ and $\sim v^2/c^2$. Below from exact relativistic
equation in the electromagnetic field $A_\mu$ we obtain this
decomposition.   Indeed,  changing the $\Psi$ function  by the
following way:
\be\label{Psi} \Psi(r,t)= \Psi(r)e^{-i(E+m)t}\ee
we obtain:
\be\label{AEV}  (E+m)\Psi(r) = \big[
\overrightarrow{\alpha}(\overrightarrow{\hat{p}}-e
\overrightarrow{A})+\beta(m_1+\gamma_5 m_2) - V\big]\Psi(r) , \ee
where $E$ energy of fermion in which non included energy of rest,
$V$-potential energy. Consider representations  of  the
four-components function $\Psi$ in a form $\Psi ={\varphi \choose
\chi }$, where $\varphi $ and $\chi$  in turn, two-component
functions and taking into account, that in standard representation
     $$\alpha\Psi={\sigma\chi \choose
     \sigma\varphi},\,\,\beta\Psi={\varphi \choose
     -\chi},\,\,\beta\gamma_5\Psi={-\chi \choose
     \varphi},$$
we can write:
 \be\label{vc1}(E+m-m_1+V)\varphi = [\vec{\sigma}\left(\vec{p} -e
 \vec{A}\right)-m_2]\chi; \ee

 \be\label{vc2}(E+m+m_1+V)\chi =[\vec{\sigma}\left(\vec{p}-e \vec{A}\right)+m_2]\varphi, \ee
  where $\vec{\sigma}$ are  matrix Pauli.  Expressing $\chi$ from
 (\ref{vc2}) we can obtain
 \be\label{vc3}\chi=\frac{\vec{\sigma}(\vec{p}-e\vec{A})+m_2}{m+m_1}\left(1+\frac{E+V}{m+m_1}
 \right)^{-1}\varphi.\ee

Taking into account that in non relativistic limit
$$
E+V \ll m+m_1,
$$
with accuracy  up to quadratic terms  on velocity of fermion $\sim
v^2/c^2$ we have

 \be\label{vc3}\chi=\frac{\vec{\sigma}(\vec{p}-e\vec{A})+m_2}{m+m_1}\varphi.\ee

Using the identity
$$
(\vec{\sigma}\vec{\hat{a}})(\vec{\sigma}\vec{\hat{b}})=\hat{\vec{a}}+i\vec{\sigma}[\hat{\vec{a}}\times\hat{\vec{b}}],
$$
 considering here $$\vec{\hat{a}}= \vec{\hat{b}}=
(\vec{\hat{p}}-e\vec{\hat{A}})^2,$$  in a result we obtain
$$
[\vec{\sigma}({\hat{\vec{p}}-e\hat{\vec{A}}})]^2=(\vec{\hat{p}}-e\vec{\hat{A})}^2-e\hat{\vec{\sigma}}{\vec{H}}.
$$
 Thus in non-relativistic limit for two-components wave function we can write

\be\label{vc4} E\varphi(r)=i \frac{\partial}{\partial t} [\varphi(r) exp(-iEt)] =
 \hat{H}\varphi(r)=\left[\frac{(\hat{\vec{p}}-e\vec{A})^2}{m+m_1}-\frac{e\vec{\sigma}\vec{H} }{m+m_1}
+V\right]\varphi(r).\ee

Thus, we see that in the obtained equations the role  mass plays
parameter
\be \label{nrm}m^* = (m+m_1)/2  \geq m.\ee
This value
essentially different from  the relativistic mass $m$ defined in
(\ref{012}). We emphasize that this distinction has nothing to do
with the value of the particles velocity.

First, contrary to the claims of \cite{Nez}, \cite{ft12}, not always Hermitian and
pseudo-Hermitian components of the mass are contained in universal
mass extension (\ref{012}) needs to correspond relativistic model
(\ref{ModDir}).

   Second, pseudo-Hermitian extensions of the usual Dirac equation
   itself  to another format  can contain not only  to some small
   differences but sometimes and  very significant values of
   relativistic and non-relativistic masses.  Here we can see that
   masses $m$ and $m^*$  is equal only in  the   limit of the
   usual Dirac's equations $m_2 \rightarrow 0$ and $m_1\rightarrow m$.

\section {An upper limit on fermion mass spectrum in
non-Hermitian models }

As already mentioned,according to elementary physical
considerations  the parameters $m,  m_1$ and $m_2$  can not be
completely arbitrary in this non-Hermitian model. In other words,
the modified Dirac equation can describe real fermions only in a
limited area of change $m, m_1$ and $m_2$. First, the positive
definiteness of the observed relativistic mass requires obvious
limit $m_2 \leq m_1$. Secondly, the mass of the fermion should
also be limited the condition $m \leq m_1$ and finally from
conditions (\ref{012}), the masses $ m_1, m_2$ and $m$ can be
directly linked with a right triangle.

At first sight seems that these boundaries are
completely sufficient that to satisfy of any questions?
 However  nobody  with help of this constraints   can't answer on
 the simple question: can under fixed parameters  $m_1$ and $m_2$
 whether exist  maximal value mass of fermions, which may be
 considered in this models?

 In fact,  the expression (6),  which is obtained from generalization of
the notion of Hermiticity and $\gamma_5$-factorization of the
Klein-Gordon's  operator is accomplished automatically for any
from right triangles constructed on this masses.  However also it
is obviously that  areas of these triangles may be found are not
equal to each other. Thus one may find the triangle with the
maximal area.   Maximal area can allow to define and the maximal
mass $M$ which we must considered as maximal possible limiting
value  of fermion mass in this model. We also see, that
non-Hermitian $\gamma_5$ mass extension in Dirac equation may be
basis to develop the  models with constraints spectrum mass of  fermions.

Really this limitation  is not arise being so obvious, but bears a
fundamental limitations in the model. If continue geometric
interpretation of the considered connection of masses, we can set
another important ratio. We are talking about comparing different
of triangles according to their square. In particular, if one
consider an area of the  two neighboring triangles forming
rectangle $S_1 = m*m_2$ its total area  may not exceed the area of
half a square, built on the hypotenuse is $S_2= {m_1}^2.$ For
visualization consider the function(see Fig.1)

\begin{figure}[h]
\vspace{-0.2cm} \centering
\includegraphics[angle=0, scale=0.4]{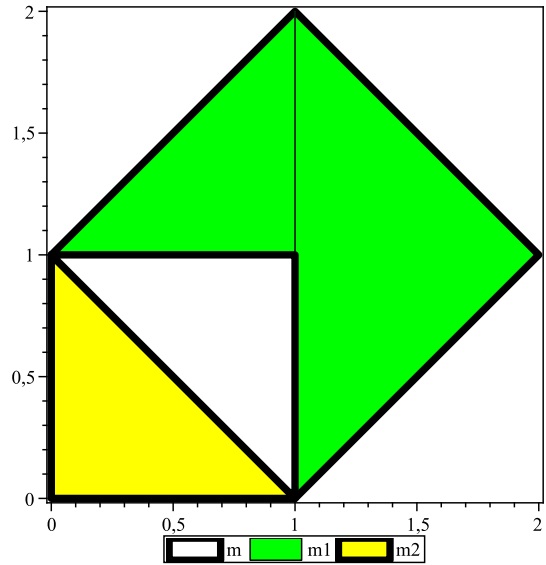}
\caption{Comparison of squares different figures constructed on
using masses $m,m_1$ and $m_2$} \vspace{-0.1cm}\label{f4}
\end{figure}

\begin{figure}[h]
\vspace{-0.2cm} \centering
\includegraphics[angle=0, scale=0.4]{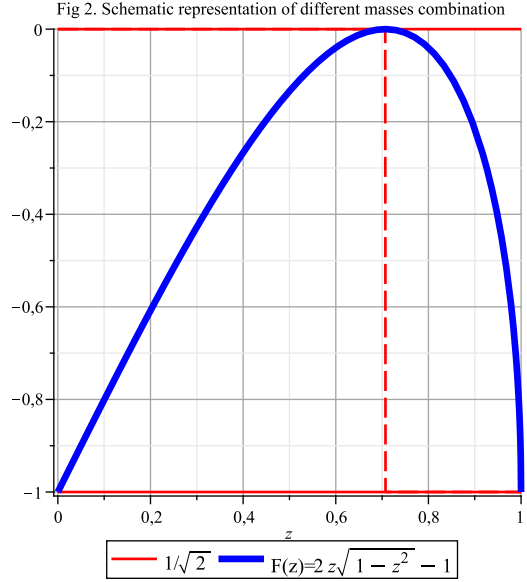}
\caption{Schematic representation of different mass combination}
\vspace{-0.1cm}
\end{figure}

However, in addition it is arise another limitation, which is not
being so obvious, but bears a fundamental limitations in the
model. Indeed, if continue geometric interpretation of the
considered connection of masses, we can set another important
ratio. We are talking about comparing different of triangles
according to their square. In particular, if one consider an area
of the  two neighboring triangles forming rectangle $S_1 = m*m_2$
its total area  may not exceed the area of half a square, built on
the hypotenuse is $S_2= {m_1}^2.$ For visualization consider the
function(see Fig.2) $$ F(m_1,m_2)= -\frac{{m_1}^2}{2}+ m*m_2\leq
0$$

From Fig.2 we can easily see that the equality sign observed in
the point $m = m_1/\sqrt{2}$. Thus, this single point, which is
establishes the equality of the areas of two squares. At the same
time it means in the model there is formed the value of the
maximum possible mass of the fermion. Denoting the maximal  mass
of the fermion in the form  $m_{max} =M={m_1}^2/2 m_2$, we now can
express the mass parameters $m_1$ and $m_2$ as  functions $M$ and
$m$ (see \cite{RodKr2}).

 Of course the question arises - to what extent  the above formulation
the definition of  maximal mass is the unambiguous definition? It
is clear that the presence in the non-Hermitian  theory of two
components of units of mass $m_1$ and $m_2$ gives our in principle
the ability to build  different variants expressions containing
$m_1$ and $m_2$.  However, it is obvious  that  making physically
reasonable choices here is in substance only one.This is result
which contains non-Hermitian mass  $m_2$ in the denominator and
itself expressions the desired type have the form
$M={m_1}^2/2m_2$. Thus, we come the expression is of the form $M
=c{ m_1}^2 /m_2$ with an arbitrary constant $c$. The value $c=1/2$
gives also the above-mentioned Cauchy's theorem on the average.

\section {The paradox of the "two masses" and its solution in
suggested model.  Exotic particles}

Let us now discuss the paradox   which  arises between two masses
in this model.  It is important that relativistic and
non-relativistic masses have absolutely different properties.  It
is inevitable when  anti-Hermitian mass $m_2 \neq 0$.  Using the
relations (\ref{888}), (\ref{nrm}) and the expression for the
maximal  mass, we can express all the components of using relation
of   relativistic  and maximal masses. In turn, we write it in the
form  $m = \nu M$. After this one may easy to see that for
non-relativistic mass  there are   two options

\be \label{nrm1}
\nu^* = \frac{m^*}{M}  =  \frac{\nu}{2} + \sqrt{ \frac{ 1\mp
(1-\nu^2)^{1/2}}{2} }  \ee

It is clear ( see expressions )  that
the upper sign here corresponds to ordinary particles and lower
sign corresponds to its exotic partners. Appropriate the graphs
shown in Fig. 3. We see that for normal particles the difference
between relativistic and non-relativistic mass is negligible.
Under such conditions, the normal particle equation (\ref{nrm1}) gives

\be \label{nrm2}  \nu^* \approx \nu + \frac{\nu^3}{16} + \frac{7 \nu^5}{256} + O(\nu^7)  \ee

This is the key point of non-Hermitian theory.
In our approach when there is  the limit of  the  parameters $m$,
$m_1$ and $m_2$, this allows to us  to solve the paradox of two
masses, when we have on one hand  the value of relativistic mass
$m$ and non-relativistic mass   $m^*$ from  the other. The
existing  of value maximal  mass in a natural way allows  to find
the  right results. Although  in the absence of this limit is the
difference between the relativistic and non-relativistic masses of
ordinary particles can to be arbitrarily large. It is definitely a
serious problem for modified theory of Dirac. In our opinion, to
find any other alternative explanation for this paradox besides
restriction of mass highly problematic.

\begin{figure}[h]
\vspace{-0.2cm} \centering
\includegraphics[angle=0, scale=0.4]{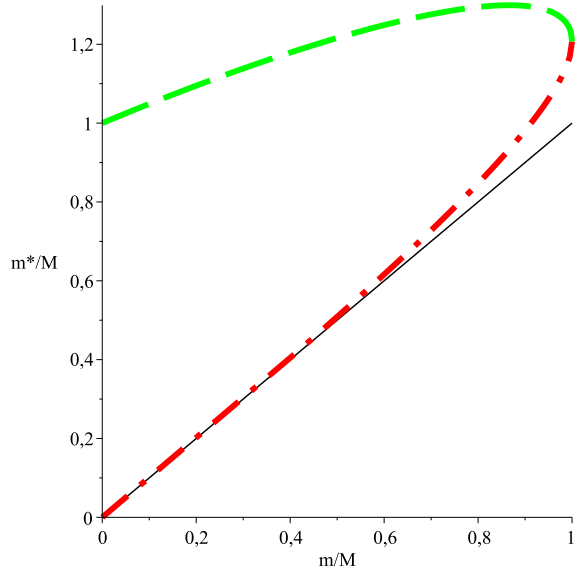}
\caption{The character of variation of the relativistic mass
(straight line) and non-relativistic mass for ordinary particles
(dash-dot line) and exotic particles (dash line).
 For the case of observed particles with small masses curves
 corresponding to non-relativistic  and  relativistic masses are practically identical. }
  \vspace{-0.1cm}
\end{figure}

Besides,  for the particles which corresponding to the upper sign
in (\ref{nrm1}) we have in our model also fairly unusual
properties. Their non-relativistic mass comparable to the maximal
masses. Such particles are extremely difficult to get to move with
acceleration, even in very strong electromagnetic fields and
therefore they  practically do not create  electromagnetic waves.
At the same time, their "gravitational charge" is determined by
their relativistic mass and hence  comparable with  masses of $m$
ordinary particles. Obviously, this masses determines the
intensity of birth and annihilation of particles. Thus, the
described particles practically do not interact with
electromagnetic waves, however   are born and are involved in
gravitational interactions the level of the ordinary particles.
All this, in our opinion, makes them the obvious candidates to the
structure of dark matter particles.

\section {The chiral representation for the modified Dirac equation. The mass spectrum of the left and right particles.
Is it possible to exist "massless" neutrinos?}

The original Dirac equation can be modified and in a slightly
different way. To this end, in he equations (\ref{D2}) and
(\ref{ModDir}) is enough to make a linear transformation

\be \label{kmass} m_{1(2)} = m_r \pm m_l  \ee
Due to the fact that

\be P_{r(l)} = (I \pm \gamma_5)/2 \ee
are the projective operators
($I$ is the identity matrix),
the Dirac equation naturally splits
into a system of two equations.
If you imagine Dirac's bispinor,
like section 3, in the form of a column of two-component
spinors
$\Psi ={\varphi \choose \chi }$ and use the standard
representation for $\gamma$-matrices, will receive

\be\label{kir1}i \frac{\partial \varphi}{\partial t} = i \vec{\sigma} \bigtriangledown \chi + (m_r+m_l)\varphi - (m_r-m_l)\chi ; \ee

 \be\label{kir2}i \frac{\partial \chi}{\partial t} = i \vec{\sigma} \bigtriangledown \varphi + (m_r-m_l)\varphi - (m_r+m_l)\chi . \ee

Acting as projective operators on the initial equation written in the form
\be i \frac{\partial \psi}{\partial t} = \left[ \overrightarrow{\alpha} \textbf{p} + 2 m_l \beta
P_r + 2 m_r \beta P_l \right] \psi \ee
or folding and subtracting the resulting equations and introducing
the notation $\xi (\eta)=\varphi \pm \chi$, come to the system

\be\label{kir3}i \frac{\partial \xi}{\partial t} - i \vec{\sigma}
\bigtriangledown \xi = 2 m_r \eta ; \ee

\be\label{kir4}i \frac{\partial \eta}{\partial t} + i \vec{\sigma} \bigtriangledown \eta = 2 m_l \xi. \ee

  The resulting equations differ from the equations Weyl  by only
  presence of  non zero     value  the "left" and "right" of the masses.
  Naturally,   the temptation   arises, using the arbitrariness of
  the initial parameters $m_1$ and $m_2$, to describe thus a massless Weyl
  particles. This is  way was used by the authors of the paper \cite{AlBen}.
  In particular, one argue that when $m_1=\pm m_2$ the system similar
  to (\ref{kir3}-\ref{kir4}) describes a massless right-wing or left neutrinos.
  In a sense it is  the limit of the original equations   of motion.

 Indeed in  this approach we are faced with quite a difficult situation.
 For example, let $m_1=m_2$. Then becomes zero not only "left mass" $m_l=m_1-m_2$,
 and also relativistic mass $m=2 \sqrt{m_l \cdot m_r}$, and the equation (\ref{kir4})
 really takes the form of a conventional two-component Weyl equation. However,
 the non-relativistic mass $m^*$ thus, generally speaking,
 in  zero is not drawn, because mass $m_1  $ can have a nonzero value. Apparently,
 to search physical meaning in such a situation under any values of masses $m_1$ and $m_2$
 is quite difficult. And really relativistic mass becomes zero  only if $m_1=m_2=0$,
 which corresponds to the well-known case of  pure Dirac neutrinos.  Note also that
 case $m_1 = m_2$ maybe realized  and for exotic particles when $m_1= m_2 =2M$.

\begin{figure}[h]
\vspace{-0.2cm} \centering
\includegraphics[angle=0, scale=0.4]{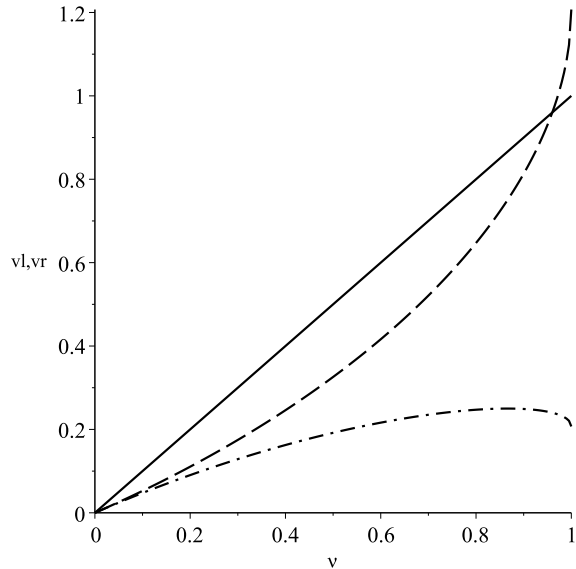}
\caption{The character of variation of the relativistic
mass (straight line), left-mass (dash-dot line) and the right-mass
(dash line)  for the case of  ordinary particles. For light particles,
the left and right masses practically coincide, which corresponds to
the usual Dirac equations.} \vspace{-0.1cm}
\end{figure}

\begin{figure}[h]
\vspace{-0.2cm} \centering
\includegraphics[angle=0, scale=0.4]{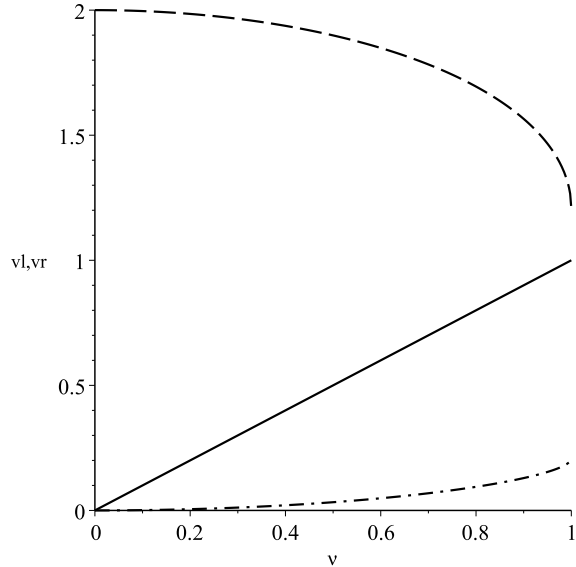}
\caption{The character of variation of the relativistic
mass(straight line), left-mass (dash-dot line) and the right-mass
(dash line)  for the case of  exotic particles. It is seen that
right-mass have  a huge values $m_r > M$, and the left-masses of
particles is very small, much smaller than the mass of the known
leptons.}
\vspace{-0.1cm}
\end{figure}

In our model with the maximal mass appearing of such a paradox in
impossible in principle. In Fig.4 shows the change of the
"starboard mass" and the "left mass" in the same scale as in
Fig.3. Appropriate expressions are obtained from the same formulas
(\ref{888}), (\ref{nrm}) and  the parameter $m = \nu M$, the same
as in the previous section:

\be \label{kirl} \nu_l = \frac{m_l}{M}  = \mu (1 - \mu) ,  \ee

\be \label{kir2} \nu_r = \frac{m_r}{M}  = \mu (1 + \mu)  , \ee
where
\be \mu = \sqrt{ \frac{ 1 \mp (1-\nu^2)^{1/2}}{2}} . \ee

The upper signs correspond to ordinary, and the lower - for
the exotic particles. For observed particles with $\nu \ll 1$, it
is easy to see that the left and right of the mass are practically
identical

\be \nu_{l(r)} \approx \nu/2 \mp \nu^2/4 , \ee
ie they correspond with great accuracy to the usual Dirac equations.
On the contrary, for exotic particles with the same
values of $\nu \ll 1$ the difference between the left and right
masses is huge:
\be \nu_l \approx \nu^2/8 \rightarrow 0; \nu_r
\approx 2-3\nu^2/8 . \ee
If, as we assume, the exotic particles
are relating to dark matter, its "background" obvious should
associate with left polarized. However, far-reaching conclusions
to do while early.

In conclusion, we note the following. The chiral representation
gives possibility to have a somewhat another understanding the
meaning of extensions of the Dirac equation. In fact, the
introduction to it additional term $\sim m_2 \gamma_5$, is
equivalent to the introduction of the difference between the right
and the left mass of the particle. But this difference is
absolutely natural element of the Standard Model (e.g. the
electro-weak interaction). The only difference is that it is now
refers to "more compact" spinor multiplet. The restriction in  the
spectrum  mass of fermions associated with a fundamental mass in a
natural way makes this difference is practically unobservable, or
giant.

\section {Conclusions and confinement}

 According to the results of the study of the
modified Dirac equation can be done the following conclusions.
Introduction pseudo-Hermitian contribution in this equation is
equivalent to the introduction of the difference between right and
left masses of the particles. Its inevitable consequence is the
paradox of differences between relativistic and non-relativistic
mass, which are different combinations Hermitian and non-Hermitian
components of the mass. If someone  want to ignore  any
restrictions for mass parameters and nevertheless are going to
continue investigation of non-Hermitian $\gamma_5$-mass extension
for fermion systems must be obtained  much more complicated
physical situation  may  arise. Particularly unpleasant situation
in which the relativistic mass of ordinary particles becomes zero
and  non-relativistic component of mass stays is not zero.

In essence, this means a departure from the principle of equivalence,
which is inevitable in such kind of theories. This is the root cause
of the difficulties described. The same time, the limitations of the
mass spectrum  by the condition $M={m_1}^2/2m_2$,
 associated with the introduction of a fundamental mass,
naturally this paradox is allowed \footnote{Moreover, this condition
makes the considered algebraic 4D-theory with a pseudo-Hermitian
extension dual geometric 5D-theory of Kadyshevsky's.}.

The ratio expressing the
observed mass through the fundamental particles, is possible in
two variants. In the first difference of the relativistic and
non-relativistic mass (as the difference between the left and
right of the masses) to the lungs (compared to maximal mass $M$)
of the particles is practically unobservable, so that the ordinary
Dirac equation is performed almost exactly. It is clear that this
variant describes the usual particles.

Moreover, it becomes possible to estimate the value of the fundamental
mass using this difference. The equivalence of the heavy and inert mass
for ordinary matter was verified experimentally. According to the last
known data \cite{Bae}, \cite{Ade}, the difference between the gravitational
and inertial masses does not exceed the value $\sim 1.4*10^{-13} $.
Associating this difference with the second term on the right-hand side (\ref{nrm2}),
we obtain $\nu \sim 1.308*10^{-4}$. Assuming a proton with a mass of 0.94 GeV as
the structural unit of the ordinary substance, we obtain a rough estimate for the
fundamental mass $M \sim 7.2TeV$. Note that the energy of this level is attainable
at the Large Hadron Collider. An experimental confirmation of our theory for
ordinary matter would be the detection of the difference in particle masses
measured from the threshold of its production, on the one hand, and its
motion in the electromagnetic field, on the other.

In the second embodiment,
on the contrary, the difference between relativistic and
non-relativistic, and the left and right of the masses is huge and
comparable to the fundamental mass. Such particles have
non-relativistic the mass of the mass of maximal, but their
relativistic mass is comparable with the mass of ordinary
particles. Therefore, they do not participate in electromagnetic
interactions, but gravity show themselves as ordinary particles.
Thus, the introduction of fundamental mass as by-product gives
clear candidates for the particles dark matter.

\end{document}